\documentclass[12pt]{article}
\usepackage[comma, super, sort&compress]{natbib}
\usepackage{tabularx}
\usepackage{graphics}
\usepackage{wrapfig}

\usepackage{graphicx}
\usepackage{braket}
\usepackage{times}
\usepackage{float}
\usepackage{booktabs} 
\usepackage{placeins}

\usepackage{bibunits}

\newcommand{\sifig}[1]{SI figure~\ref{sifig:#1}}

\usepackage[font={normalsize, it},labelfont=bf]{caption}

\usepackage[colorinlistoftodos]{todonotes}

\usepackage{siunitx}
\DeclareSIUnit{\pH}{pH}
\DeclareSIUnit{\Torr}{Torr}
\DeclareSIUnit{\electrons}{electrons}
\DeclareSIUnit{\gramcat}{g_{cat}}
\DeclareSIUnit{\atomicpercent}{at.\%}
\DeclareSIUnit{\gram}{g}
\DeclareSIUnit{\percent}{\%}
\DeclareSIUnit{\sccm}{sccm}
\DeclareSIUnit{\electronvolts}{eV}
\DeclareSIUnit{\ppm}{ppm}
\DeclareSIUnit{\ppb}{ppb}
\DeclareSIUnit{\grating}{gr}
\DeclareSIUnit{\gauss}{gauss}
\DeclareSIUnit{\rpm}{rpm} %radian per minute

\usepackage{comment}
\usepackage{miller}

\usepackage{xcolor}
\usepackage{soul}

\usepackage{chemformula}
\setchemformula{kroeger-vink=true}

\usepackage{hyperref}
\hypersetup{
    colorlinks=true,
    linkcolor=black,
    filecolor=black,      
    urlcolor=black,
    citecolor=black,
}
\begin{document}

% ---------------------------
% Generic title page (edit)
% ---------------------------
\begin{titlepage}
  \centering
  \vspace*{0.5cm}

  % Title (large)
  {\large\bfseries Delta-Doped \ch{GeV-} in diamond with isotope resolvable optical spectra}
  \vspace{0.5cm}

  % Authors
  {Ian N. Hammock\textsuperscript{1},
  Kinfung Ngan\textsuperscript{2},

Antoni W. Wellisz \textsuperscript{1}, 
Keira Linder \textsuperscript{1}, 
 % Tanvi Deshmukh\textsuperscript{3},
  Midrel Wilfried Ngandeu Ngambou\textsuperscript{3,4},
Jiefei Zhang\textsuperscript{1,3,4}, 

Xinghan Guo\textsuperscript{1},
Benjamin Pingault\textsuperscript{1,3,4}, 
  Shuo Sun\textsuperscript{2},
    David D. Awschalom\textsuperscript{1,3,5},
    F. Joseph Heremans\textsuperscript{1,3,4},
  Alexander A. High\textsuperscript{1,3},
  Nazar Delegan\textsuperscript{1,3,4,*}
  \par}

  \vspace{0.5cm}

  % Affiliations

  \textsuperscript{1}Pritzker School of Molecular Engineering, University of Chicago, Chicago, Illinois 60637, USA\\

\textsuperscript{2} JILA and Department of Physics, University of Colorado Boulder, Colorado 80309, USA\\

\textsuperscript{3}Materials Science Division and Center for Molecular Engineering, Argonne National Laboratory, Lemont, Illinois 60439, USA\\

    \textsuperscript{4}Q-NEXT, Argonne National Laboratory, Lemont, Illinois 60439, USA\\

\textsuperscript{5}Department of Physics, University of Chicago, Chicago, Illinois 60637, USA\\
  \vspace{0.5cm}

  % Corresponding author
\textsuperscript{*}Corresponding author: \href{mailto:ndelegan@anl.gov}{ndelegan@anl.gov}\\
% \textsuperscript{o}Contributed equally to this work.

  \vspace{0.5cm}

  % Optional logo (uncomment and supply file)
  % \includegraphics[height=2.0cm]{logo.pdf}\\[1.0cm]

  % Optional short abstract on title page
  \begin{minipage}{0.85\textwidth}
    \small\noindent\textbf{Abstract}

    Group IV color centers in diamond  -- silicon, germanium, tin, and lead-vacancy centers -- are a promising platform for quantum networking. Typical Group-IV creation utilizes ion implantation during diamond growth that leaves the color centers with a broad depth distribution and damage-induced strain inhomogeneity. Deterministic placement and damage reduction in Group IV generation are essential for improving this platform. In this letter, we report on the localization of germanium-vacancy (\ch{GeV-}) to a $<$ 10 nm layer via the timed injection of $\text{GeH}_4$ during microwave plasma chemical vapor deposition. In this layer we have realized 17 ppm of Ge and a strain-free environment for the \ch{GeV-}. Subsequent electron irradiation and annealing have generated a sheet density of 0.12 \ch{GeV-}/$\mu$m$^2$. The minimal damage from color center generation mitigates spectral inhomogeneity and allows spectral resolution of isotopic clusters, with the best cluster having an inhomogeneous distribution of 2.14 GHz and an average linewidth of $ 47 \pm 11$ MHz. Our study gives insight into formation dynamics of \ch{GeV-} in delta-doped systems and estimate our vacancy to \ch{GeV-} yield to be $\geq~90\%$, establishing an athermal nearest-neighbor formation model. We further characterized the spin $T_2^*$ and set a lower bound of $730~\text{ns}$, limited by a paramagnetic defect.

  \end{minipage}

  \vfill

  % Keywords and date
  \vspace{0.6cm}
  \small\textbf{Keywords:} diamond, spin defects, processing, CVD, synthesis, annealing.\\[6pt]
  \normalsize \today

\end{titlepage}

% Start main text on page 1
\setcounter{page}{1}

\clearpage

\begin{bibunit}[Science]
  \section{Introduction}
Group IV color centers in diamond have coherent spin and optical interfaces that are utilized in quantum network architectures and in investigations of many-body physics. These color centers are insensitive to first-order electric field noise and have demonstrated excellent single qubit properties in nanophotonic and strained devices \cite{meesala_strain_2018, machielse_quantum_2019,guo_microwave-based_2023, ding_high-q_2024}. Moreover, entangled two qubit and blind computing has been demonstrated with Group IVs, enabled by quantum frequency conversion \cite{bersin_telecom_2024, knaut_entanglement_2024, wei_universal_2025}. In parallel, heterogeneous integration with $\chi^{(2)}$ materials such as lithium niobate has been used to demonstrate phononic and photonic control and explore transduction schemes \cite{xu_thin_2025, ding_heterogeneously_2026}. More recently, two-dimensional arrays of Group-IV centers in diamond were investigated as a platform to engineer compact optics\cite{andreoli_maximum_2021, andreoli_metalens_2025}, and could be extended to enable emergent cooperative effects \cite{asenjo-garcia_exponential_2017}, or squeezed spin\cite{hosten_measurement_2016, gao_signal_2025} in large identical ensembles. 

Current synthesis typically relies on ion implantation, where dopant-lattice collisions generate Frenkel pairs and strain-inducing vacancy clusters. Due to the large strain susceptibility of group IV color center, this heterogeneity causes significant spectral inhomogeneity \cite{iwasaki_germanium-vacancy_2015, meesala_strain_2018} and other electrostatic issues\cite{hammock_diamond_2026}. Additionally, for device integration, deterministic spatial localization is key for achieving desired performance \cite{rugar_quantum_2021}. However, implanted ions present an inherent straggle on the order of tens-of-nm (scaling with the implantation energy) \cite{rugar_quantum_2021,schroder_scalable_2017,froch_versatile_2020}. High pressure high temperature (HPHT) annealing techniques, in which the diamond lattice is brought to thermodynamic regimes that promote sp$^3$ stability, can reduce inhomogeneity to desireably gigahertz (GHz) levels \cite{narita_multiple_2023, harris_coherence_2024,baba_annealing_2025, abe_narrow_2026} but does not address depth localization. Shallow implantation with subsequent overgrowth, demonstrated to generate SnV ensembles with a straggle under 3 nm \cite{rugar_generation_2020}, can provide excellent depth localization but with significant 100 GHz inhomogeneity. 

Alternatively, growth-based doping can yield precise control over depth localization while sidestepping ion-implantation-induced lattice damage. Growth-based doping methods such as HPHT\cite{ekimov_germaniumvacancy_2015, palyanov_germanium_2015, palyanov_high-pressure_2016, westerhausen_controlled_2020} and chemical vapor deposition (CVD)\cite{rogers_electronic_2014, iwasaki_germanium-vacancy_2015, zhang_micron-sized_2019, grudinkin_photoluminescence_2020,sedov_photoluminescence_2014, sedov_growth_2018, sedov_formation_2022, sedov_microporous_2023, yurov_situ_2023,pavlenko_low-temperature_2024} have been used to generate optically homogeneous Group-IV color centers \cite{sipahigil_indistinguishable_2014} with spin properties that surpass implantation-derived results. By further leveraging intrinsic control over precursor residency times and slow growth rates, depth-localization of color centers (i.e. $\delta$-doping\cite{ohno_engineering_2012}) with accuracies of $\approx$1.6 nm was shown to be possible\cite{wyatt_creation_2025,marcks_quantum_2024,rogers_electronic_2014, iwasaki_germanium-vacancy_2015, sedov_formation_2022}.

In this work, we report on the in-situ synthesis of delta-doped germanium vacancy centers in isotopically purified diamond. We investigate in-situ germanium incorporation and depth localization, followed by subsequent conversion into GeV$^-$ centers through post-processing.  Our approach allows us to decouple key formation pathways by quantitatively studying vacancy conversion efficiencies, providing insight into defect formation dynamics. Photolouminescence spectra of the GeV$^-$ reveal minimal strain in the doped layer and isotopic resolution of the GeV$^-$ zero phonon lines. We observe excellent optical homogeneity within each isotopic grouping. Finally, all-optical measurements of spin coherence reveal coherence times comparable to implantation-created GeV, likely limited by an unknown paramagnetic species.

\section{Diamond Growth and Processing}
\begin{figure}[H]
    \centering
    \includegraphics[]{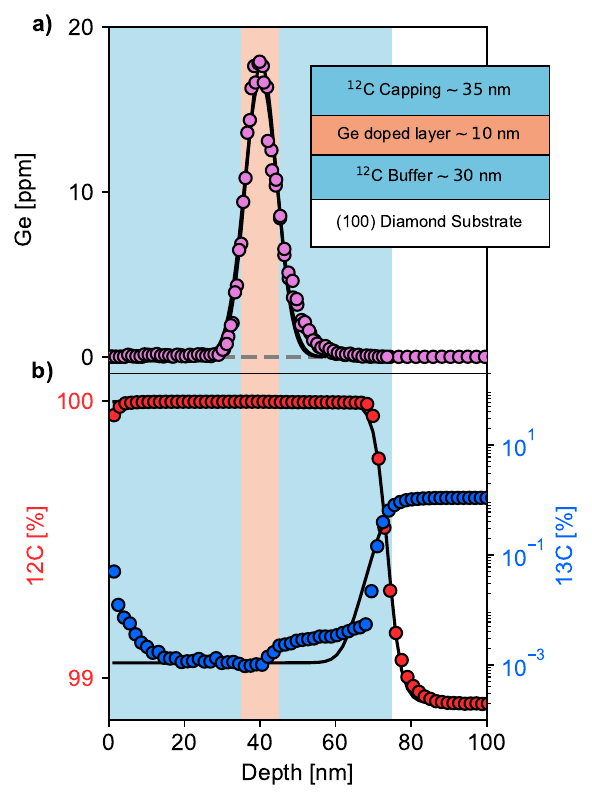}
    %The trim order is left bottom right top.
    \caption{Secondary Ion Mass Spectroscopy (SIMS) of a) Germanium and b) Carbon Isotopes: 12 and 13 as a function of depth. Detection limit of Ge is given as the dashed line at 4 ppb.  a inset) A schematic illustrating the diamond overgrowth stack with approximate thicknesses.}
    \label{fig:fig1}
\end{figure}

Diamond overgrowth was performed on a strain released \cite{guo_tunable_2021} 2x2x0.5mm substrate (EL grade, ElementSix) in a microwave plasma chemical vapor deposition (MPCVD) chamber, using a  $\geq 99.95~ \text{at.}\%$ iso-purified $^{12}\text{CH}_4$ carbon precursor with a 0.1:100 ratio of $^{12}\text{CH}_4$:$\text{H}_2$ ($\approx$14.2 nm/h growth rate). Growth was performed at an on-substrate temperature in the range of 800-875 $^\circ$C as measured via optical pyrometer and a plasma power density of $\approx$24 W/cm$^3$, at a pressure of $25$ Torr. Ge doping was achieved by introducing diluted germane, with a total dilution of $\text{GeH}_4$:$\text{H}_2$ of 1.75:1000. The sample geometry is schematized in the inset of figure \ref{fig:fig1}b, with the associated Secondary Ion Mass Spectroscopy (SIMS) depth profile showed in figure \ref{fig:fig1}a and b, for the Ge and $^{12}$C:$^{13}$C ratios, respectively. After growth, the substrate was degassed at $\approx$ 550 $^\circ C$   for 14 hours in high-vacuum ($\leq 2\times10^{-7}$ Torr) to minimize undesirable hydrogen related defect formation in subsequent processing steps\cite{hammock_diamond_2026}. 

SIMS characterization confirms at 40 nm deep, a $17$ ppm Ge layer with a $9$ nm FWHM, with an achievable lower bound based on precursor residency times of around 2 nm\cite{marcks_quantum_2024}.Integrating over the layer yields a sheet density of $33,000$  Ge/$\mu \text{m}^2$. To achieve a similar amount of germanium incorporation with ion implantation at this depth would require a dose of around $10^{13}$ ions/cm$^2$, but the distribution\cite{ziegler_srim_2010} would be over $20$ nm and create a peak density of vacancies around $10,000$ ppm, see \sifig{SRIM}. This vacancy density is approaching diamond's amorphization threshold of $50,000$ ppm\cite{fairchild_fabrication_2008} and would cause extreme inhomogeneity within the diamond. No Nitrogen was measured in the sample to a detection limit of 17 ppb, likewise with no oxygen down to the detection limit of $0.6$ ppm, see \sifig{SI_sims_NO}. Moreover, the isotope of $^{13}$C at the Ge layer is measured to be on the order of $10^{-3}~\%$ - down two to three orders of magnitude from natural abundance, see figure \ref{fig:fig1}b. We observe an irregularity in  $^{13}$C incorporation in the first $\approx$ 15 nm of growth. This is being investigated and it remains unclear if this arises as a SIMS artifact (re-sputtering of natural abundance) or is an actually true growth anomaly (e.g. natural abundance carbon contamination). Critically, this anomaly stops at the Ge doping layer.

Defect activation is achieved with electron irradiation at 2 MeV with a dose of $2\times10^{14}$ e$^-/\text{cm}^2$, resulting in an expected vacancy density of $\approx$ 0.25 ppm\cite{campbell_radiation_2000, schwartz_effects_2012}. A subsequent vacuum annealing regimen is then performed at 850 $^\circ$C and 1400 $^\circ$C \cite{hammock_diamond_2026}.

\section{Formation Dynamics}
Unlike ion implanted samples, Ge atoms are incorporated during growth predominantly to the substitutional site, due to the nature of CVD thermodynamics \cite{goss_vacancy-impurity_2005}. We intentionally maintain a low growth rate \cite{rogers_electronic_2014, iwasaki_germanium-vacancy_2015} to maximize crystal quality and minimize the in-situ formation of vacancies and other undesirable species. Therefore, ex-situ vacancy creation is necessary to generate GeVs. Using electron irradiation and subsequent annealing, we generate a 0.12 \ch{GeV-}/$\mu\text{m}^2$ sheet density, see figure \ref{fig:fig2}a. We consider two parallel mechanisms for \ch{GeV-} creation\cite{hammock_diamond_2026}: the ejection of a carbon adjacent to the germanium substitutional site, an athermal process, and/or delocalized vacancy creation and migration to the germanium site. In ion implantation studies with Sn\cite{wahl_direct_2020} and Ge\cite{wahl_structural_2024}, beta decay is experimentally used to suggest the ejection of a neighboring site is likely. Annealing serves only to clean the lattice via migration of vacancies and easily disassociated clusters but overall hinders XV formation (where X is a generic substitutional dopant larger than the C atom). Other NV studies align with this and suggest that the adjacent C-atom ejection is the energetically favorable, and therefore most likely route\cite{deak_formation_2014} compared to the migrating and combining vacancy picture. However, NV studies using spatially localized vacancies\cite{onoda_diffusion_2017} suggest that migration could still be the dominant mechanism under other circumstances.

At annealing temperatures of $\geq 850~^\circ\text{C}$, vacancy diffusivity reaches $\approx 2\text{ nm}^2/\text{s}$\cite{ohno_three-dimensional_2014}, driving competing defect recombination, surface loss, and trapping processes \cite{newton_recombination-enhanced_2002, iakoubovskii_annealing_2003}. Mobile vacancies recombine with carbon self-interstitials \cite{prins_annihilation_2000, newton_recombination-enhanced_2002, cheng_laser_2025} and annihilate at surfaces and interfaces, creating a vacancy-depleted layer near the sample surface or any other significant interface \cite{racke_vacancy_2021}. Concurrently, vacancies can be captured by substitutional impurities (e.g. N, Ge) or secondary vacancies to form $X\text{V}$ color centers and multivacancy complexes \cite{goss_vacancy-impurity_2005, slepetz_divacancies_2014, gali_first-principles_2026}. At high local vacancy concentrations, mobile divacancies (V$_2$) can also migrate directly to Ge impurities to form $\text{GeV}$ centers \cite{baba_annealing_2025}. 

Here, the \ch{GeV-} yield remains invariant between the $850~^\circ\text{C}$ and $1400~^\circ\text{C}$ anneals (\sifig{SI_Anneal}). While the emergence of \ch{GeV-} centers after $850~^\circ\text{C}$ does not inherently distinguish athermal creation during irradiation from thermally activated vacancy mobilization/capture, the yield saturation by $850~^\circ\text{C}$ indicates the governing formation process is complete at or below this temperature. This temperature invariance establishes two main considerations on the vacancy kinetics: (1) divacancy (V$_2$) migration to Ge impurities is negligible in this regime \cite{baba_annealing_2025}, and (2) the mobile monovacancy pool is fully exhausted by $850~^\circ\text{C}$. Furthermore, the thermal stability of \ch{GeV-} centers up to $1400^\circ\text{C}$ agrees with their high calculated dissociation energy \cite{goss_vacancy-impurity_2005, gali_first-principles_2026}. The absence of \ch{GeV-} annihilation at elevated temperatures—despite exceeding the thermal activation barrier for carbon self-interstitial mobility—further points to a negligible residual interstitial concentration \cite{hammock_diamond_2026, gali_first-principles_2026}. 

These results demonstrate that $\text{GeV}$ creation is predominantly driven by athermal processes, specifically the direct displacement of a carbon atom neighboring a substitutional germanium site ($\text{Ge}_\text{s}$) during electron irradiation. For a $\text{Ge}_\text{s}$ center coordinated by four nearest-neighbor (NN) carbon atoms, each carbon has an ejection probability of $\ge 0.25\text{ ppm}$, for this irradiation\cite{campbell_radiation_2000,schwartz_effects_2012,deak_formation_2014}, facilitated by local strain and $\text{Ge}-\text{C}$ bond weakening relative to bulk sp$^3$ $\text{C}-\text{C}$ bonds. Summing over the four NN sites yields a direct vacancy-generation probability of $\geq ~ 1.0\text{ ppm}$ at the $\text{Ge}_\text{s}$ atom, corresponding to an estimated $\text{GeV}$ density of $0.033\text{ GeV}/\mu\text{m}^2$. 
Following DFT arguments, vacancies created at the 12 next-nearest-neighbor (NNN) sites likely spontaneously capture into $\text{Ge}_\text{s}$ due to local strain field gradients \cite{kuate_defo_how_2019, gali_first-principles_2026}, and as such remain an athermal process. Summing over all 16 NN and NNN candidate sites increases the predicted yield to $\approx 0.13\text{ GeV}^-/\mu\text{m}^2$, in agreement with our experimentally observed 0.12 \ch{GeV-}$/\mu\text{m}^2$ (Figure~\ref{fig:fig2}). Although only a small percentage, $36 ~\times~10^{-5}\% $, of Ge are converted to \ch{GeV-}, 90\% of NN and NNN vacancies are converted to \ch{GeV-}. 

Computational arguments suggest that the unavoidable contamination of nitrogen donors will likely incorporate near group IV dopants during growth \cite{chuprina_photoelectric_2026}. This is advantageous, as the nearby donor likely stabilizes the charge of \ch{GeV} to the negative state. In this context, we propose that annealing at $\ge 850~^\circ\text{C}$ is critical for mobilizing and removing residual, non-complexed vacancies, which would otherwise act as competing electron traps that would destabilize this charge state. However, dense ensemble creation of \ch{GeV-} will still require deliberate Fermi-level engineering \cite{luhmann_coulomb-driven_2019, luhmann_charge-assisted_2021}. Systematic dose-dependent irradiation studies across a broader thermal activation range are required to further decouple secondary defect pathways.

\section{Optical}

\begin{figure}[H]
    \centering
    \includegraphics[]{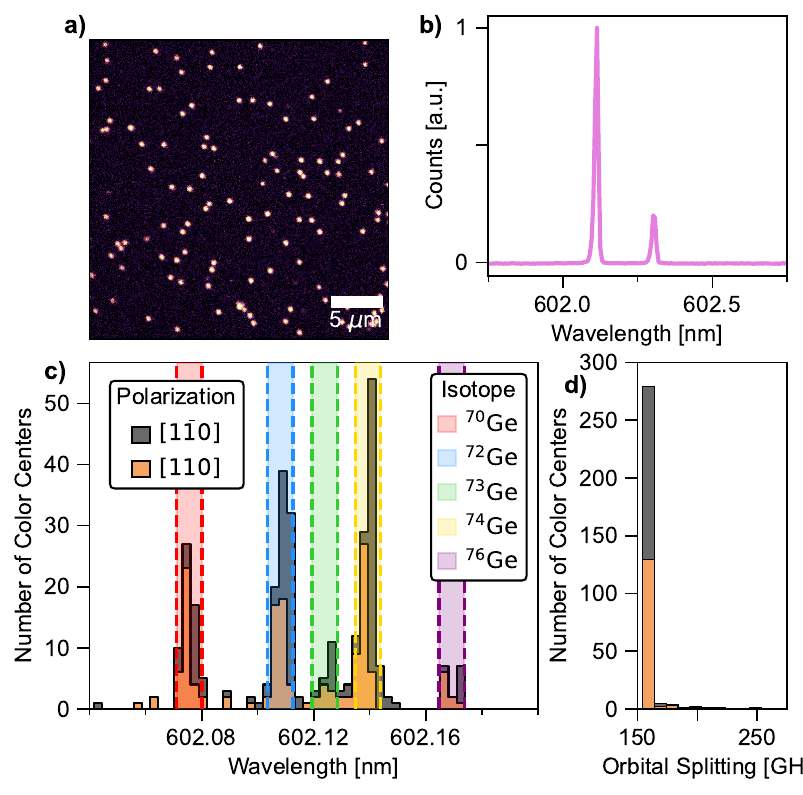}
    %The trim order is left bottom right top.
    \caption{a) PL map of the diamond with a narrowband filter in place. b) Representative spectra of the GeV$^{-}$. The left most peak is referred to as the ZPL, the distance between the two peaks is the orbital splitting. c) Distribution of GeV$^-$ ZPLs resolved by the spectrometer, distinguished by polarization. The two polarizations are stacked for clarity. Proposed isotopic grouping is overlayed. Group width is estimated by the $^{76}$Ge cluster and is 7.5 GHz wide d) Distribution of orbital splitting, distinguished by polarization.}
    \label{fig:fig2}
\end{figure}

Using off resonant excitation, we measure GeV$^-$ photoluminescence spectra, shown in Figure \ref{fig:fig2}b. From the spectra, we extract the zero-phonon line (ZPL) and generate a histogram of five distinct clusters that are agnostic of polarization. We hypothesize these clusters to originate from the Germane gas' isotope distribution causing a shift in the emission. This emission can be understood as a change in the vibronic structure due to the different Ge masses\cite{dietrich_isotopically_2014}, see table \ref{tab:iso}. We identify the bluest cluster to be $^{70}$Ge, and the reddest $^{76}$Ge. We then use the $^{70}$Ge and $^{76}$Ge clusters to estimate where $^{72}$Ge, $^{73}$Ge, and $^{74}$Ge clusters should be from their expected shifts to the optical spectra. This estimate and subsequent population quantification is portrayed in figure \ref{fig:fig2}c. The width of the shaded boxes (7.5 GHz)  are identical and bounded by the observed and isolated $^{76}$Ge cluster. We approximate that all of the populations will have similar inhomogeneous broadening effects. From figure \ref{fig:fig2}c we can see that the proposed boxes for $^{72}$Ge, $^{73}$Ge, and $^{74}$Ge are in good agreement with the observed clustering. Furthermore, we observe that the trend matches well with the expected natural abundance of the precursor gases, see table \ref{tab:iso}. We note that there is likely overlap between the $^{72-74}$Ge distributions, making it a challenge to unambiguously assign each isotope. Color centers that are not unambiguously within a cluster are marked as $^{??}$Ge. 

Additionally, $87\%$ of the observed color centers exhibit an orbital splitting of $160 \pm 8\text{ GHz}$, closely matching the theoretical unstrained value for $\text{GeV}$ centers \cite{thiering_ab_2018} (Figure~\ref{fig:fig2}d). Because defect centers occupying different crystallographic orientations sample distinct components of the local strain tensor \cite{meesala_strain_2018}, this uniform distribution confirms negligible strain within the \ch{GeV-} doped layer. Comparing the orbital splitting and ZPL histograms reveal that in this low-strain regime, inhomogeneous broadening is sufficiently suppressed such that secondary effects from isotopic variation, to govern the observed ZPL distribution.

\begin{figure}[H]
    \centering
    \includegraphics[]{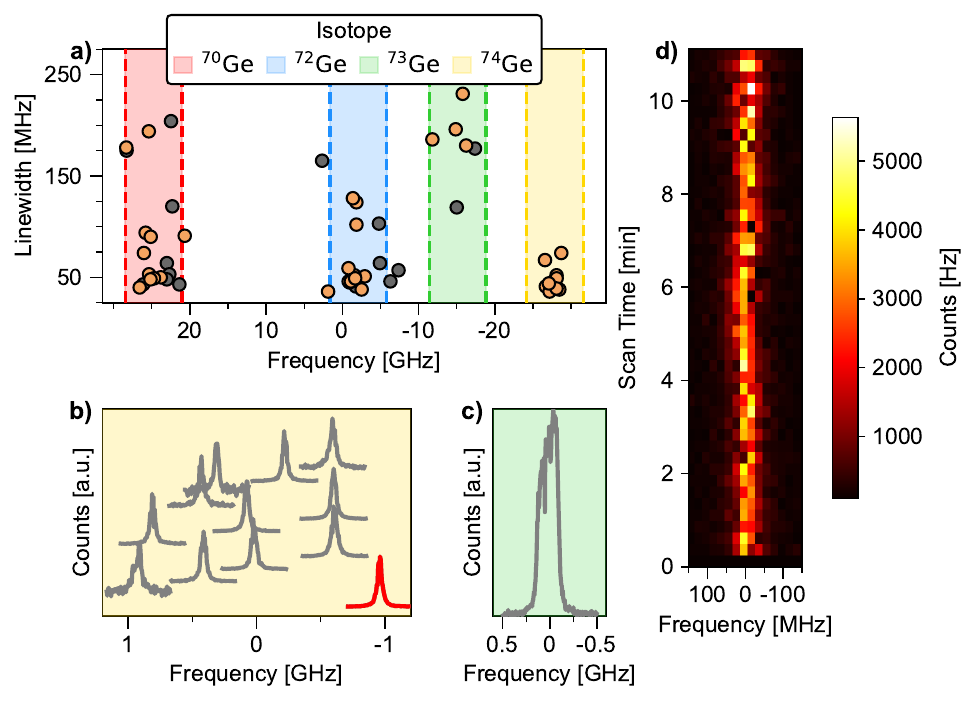}
    %The trim order is left bottom right top.
    \caption{a) Distribution of GeV$^-$ ZPLs resolved by PLE, distinguished by polarization. Proposed istopic grouping is overlayed. The span of each grouping is 7.5 GHz and determined from the $^{76}$Ge Spectra.  b) PLE scans of the $^{74}$Ge group. c) PLE scan of a $^{73}$Ge. d) 10 minute PLE scan of the highlighted color center in figure \ref{fig:fig3}b, showcasing the optical stability under resonant excitation. }
    \label{fig:fig3}
\end{figure}

To further resolve the optical spectra and isotopic sub-populations, we perform photoluminescent excitation (PLE), to investigate with MHz resolution the linewidth and ZPL. In these experiments, we keep the resonant power around 2$-$5 nW to avoid power broadening effects, and scan over a 1.5 GHz range for 10 minutes. Representative color centers, from the 322 observed, were randomly down-selected from each cluster due to experimental time constraints, with PLE measured successfully with a ratio of 70:13. Of these 70 individual color centers: about 49 of them have stable PLE with 1 exhibiting stable emission but two distinct peaks, see \sifig{SI_PLE_101}. In the remaining 20: 5 of them required resonant excitation greater than 10 nW, 9 were slightly unstable (see SI) and 6 had linewidths greater than 400 MHz. We attribute the instability to a nearby defect, either a divacancy or nearby other color centers (e.g. NV), and hypothesize that a gated PLE scheme could reduce this\cite{gorlitz_coherence_2022}. In the following analysis, we exclude the 6 broad and the 5 high resonant power color centers as their signals were immeasurable in typical experimental conditions. A representative stable GeV is given in figure \ref{fig:fig3}d, showing no signs of hopping or spectral instability\cite{ li_atomic_2024, pieplow_efficient_2024}. However, stable color centers still undergo self ionization\cite{wang_transform-limited_2024}, but recover immediately, see SI. 

In figure \ref{fig:fig3}a, we plot the linewidth versus ZPL for each color center, while  maintaining the same groupings as figure \ref{fig:fig2}. We find that within each cluster the ZPLs are distributed within $<$ 10 GHz of each other, see table \ref{tab:iso_summary}. This is within range of demonstrated strain or stark tuning devices to tune zero-phonon line emission and realize identicality \cite{meesala_strain_2018, machielse_quantum_2019, bushmakin_two-photon_2025,brevoord_large-range_2025}. Additionally, the $^{74}$Ge group hosts excellent color centers with the measured color centers being within 2 GHz of each, and an average linewidth of 47 MHz; this sub-group of PLE is plotted in figure \ref{fig:fig3}b. In contrast, the $^{73}$Ge group has broader PLE, illustrated in figure \ref{fig:fig3}c. This is in agreement with the zerofield splitting of the spin 9/2 system of the $^{73}$Ge nucleus\cite{harris_hyperfine_2023}. Excluding this "naturally" broadened $^{73}$Ge contribution, our doping scheme results the linewidth average to be 73 MHz, and around 50\% of the measured color centers to have linewidths less than 50 MHz (3.12 ns) see figure \ref{fig:fig3}a, demonstrating that nonradiative decay is minimized.

\section{Spin}
Focusing on characterizing the electronic spin of our delta-doped GeVs, we employ coherent population trapping (CPT)\cite{rogers_all-optical_2014, pingault_all-optical_2014}, an all optical technique that creates a lambda system to address the GeV$^-$ spin directly. A schematic of group IV electronics and the lamba scheme is given in \sifig{SI_lamb}. Here, all optical control is advantageous to microwave control because the low transverse strain prevents the qubit levels from exhibiting a large spatial overlap\cite{debroux_quantum_2021, guo_microwave-based_2023}. In addition, the low transverse strain enables decoherence pathways between the orbital manifolds, therefore milikelvin temperature operation is critical to suppress phononic decoherence pathways \cite{siyushev_optical_2017, sohn_controlling_2018}. 

\begin{figure}[H]
    \centering
    \includegraphics[]{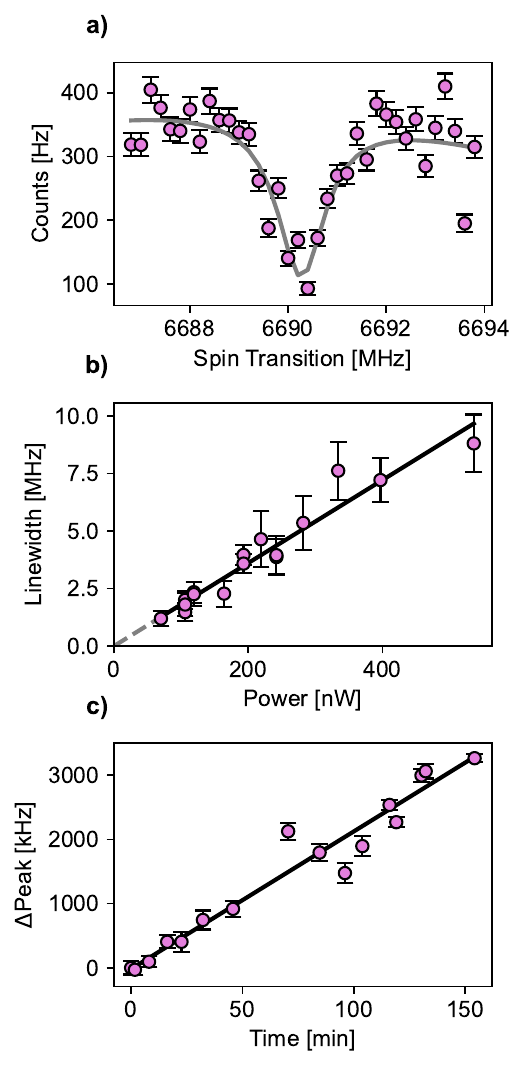}
    %The trim order is left bottom right top.
    \caption{a) CPT dip for GeV$^-$ at our lowest power, 70 nW. Fitted line is given in grey. See SI for further fit details. b) Change of CPT linewidth with incident laser power. Fitted line is given in black, with the extrapolation to 0 power in grey. c) Change of the CPT resonance over time.}
    \label{fig:fig4}
\end{figure}

Using this lambda scheme, we demonstrate a characteristic CPT dip and use a quadratic model to fit the PLE background and a Lorenztian for the dip (see SI). At $70$ nW we extract a linewidth of 1195 $\pm ~334$ kHz ($T_2^*$ = $133\pm37~\text{ns}$), see figure \ref{fig:fig4}a. Furthermore, we repeat this process across many optical powers to determine the linewidth as a function of power. This broadening will depend linearly as a function of power, and fitting it to a linear model sets a lower bound to the $T_2^*$. The fitted model extracts an intercept of $0 \pm  218 ~ \text{kHz}$ with an $R^2= 0.93$. From this we set the lower bound of $T_2^*$ to be $730 ~\text{ns}$, see figure \ref{fig:fig4}b. This value is lower than anticipated compared to previous CPT measurements under similar conditions ($T_2^* = 1.43~\mu\text{s}$ \cite{senkalla_germanium_2024}) or strained $\text{SnV}^-$ at 4 K ($T_2^* = 2.7~\mu\text{s}$ \cite{guo_microwave-based_2023}). We attribute this reduction in $T_2^*$ to spectral diffusion caused by a nearby charging paramagnetic defect, which induces a temporal drift in the spin resonance frequency (Fig.~\ref{fig:fig4}c).  Off-resonant photoexcitation with a green or blue laser resets this frequency shift to zero, after which the resonance climbs steadily at a rate of $\sim$20 \text{kHz/min} (\sifig{SI_CPTChange}). To understand this defect we employed electron spin resonance and found a weak g factor of around 2 within the diamond, see \sifig{SI_EPR} . This measurement did not help elucidate the source of paramagnetism, as many defects in diamond have a similar g factor \cite{lomer_electron_1973,iakoubovskii_vacancy_2004, baker_deducing_2007} (details in SI). Subsequent experiments are planned to elucidate the origin of this defect.

Additionally, future studies will move away from bulk material into a diamond membrane platform \cite{guo_tunable_2021, guo_direct-bonded_2024}, enabling longer $T_2^*$ coherence times for single spin qubits through the application of transverse strain \cite{sohn_controlling_2018, guo_microwave-based_2023} while giving us more latitude to isolate the cause of this paramagnetic defect species.

\section{Conclusion}
In summary, we have demonstrated the synthesis of low-strain delta-doped germanium in diamond via chemical vapor deposition (CVD). SIMS measurements confirm a high Ge incorporation density approaching the kinetic solubility limit of $< 60$ ppm\cite{iwasaki_germanium-vacancy_2015, mary_joy_photoluminescence_2024}. Optical spectroscopy of the resulting germanium-vacancy (\ch{GeV-}) centers confirms a low-strain environment enabling isotope-resolved optical transitions and narrow linewidths. In addition, electron irradiation was used to propose a refinement on their formation mechanisms in the diamond lattice. However, these results are observed alongside an unidentified paramagnetic defect that warrants further investigations. This growth process is readily compatible with thin diamond membranes \cite{guo_tunable_2021, guo_direct-bonded_2024}, establishing a scalable platform for strain-engineered quantum devices, integrated photonic/phononic resonators, and heterogeneous integration with nonlinear optical media.

\section{Methods}
\subsection{Optical Experiments}
Optical measurements were taken at 4 K, in a Montana CryoAdvance 100 with a 0.9 NA cryo-objective. Color centers were addressed with a home-built confocal microscope. Off resonant photoluminesence measures used a 532 nm laser with around 3 mW of power. PL maps were obtained using a tilted narrowband filter (Alluxa 607.7-1.6). Color center spectra was characterized via a Horiba IHR550 spectrometer, equipped with a 2400 g/mm grating and a Synapse Plus 2048 Camera. For PLE, we used an M Squared SolsTIS laser and sum frequency generation (AdvR) \cite{guo_tunable_2021, li_atomic_2024} to obtain the orange light. The phonon sideband was collected with a Semrock 647/57 bandpass filter and sent to an Excelitas avalanche photodiode. Measurement software was developed using the nspyre\cite{feder_nspyre-orgnspyre_2023} package.

\subsection{Spin Measurements}
The sample was mounted in a Bluefors dilution refrigerator (LD400) and cooled down to 20 mK. Optical excitation and signal collection were carried out by a home-built confocal microscope with a high NA objective (Olympus MPLFLN100x, NA = 0.9, working distance = 1 mm). The resonant excitation laser was generated by sum-frequency generation (SFG) of 1550 nm (Toptica CTL1550) and 984 nm (Moglabs LDX) lasers in a Periodically Poled Lithium Niobate (PPLN) crystal (HCP photonics). The 1550 nm laser was first sent through an amplitude Electro-Optic Modulator (EOM, Exail) and subsequently amplified by an erbium-doped fiber amplifier (Thorlabs EDFA100P) before entering the PPLN. The EOM was driven by a microwave source (Windfreak SynthHD v2) at the frequency of the GeV$^-$ electron-spin resonance, resulting in the two phase coherent sidebands that were used for coherent population trapping. The 602 nm, 445 nm, and 532 nm lasers pulses were generated using acousto-optic modulator (AA Optoelectronic) in their respective optical paths, with pulse timing controlled by a digital signal generator (Swabian Instruments). The phonon-sideband emission is collected and measured with a single-photon counting module (Excelitas), and the resulting electrical pulses were recorded by a time-tagging module (Swabian Instruments).

\section{Figures and Tables}

\begin{table}[ht]
    \centering
    \caption{Measured Abundance of each isotope}
    \label{tab:iso}
    \begin{tabular}{c c c}
        \hline
        Isotope & Natural Abundance & Measured Abundance (N = 322)  \\
        \hline
        $^{70}$Ge & 20.5$\%$ & 18.01$\%$ (58)\\
        $^{72}$Ge & 27.4$\%$ & 27.02$\%$ (87)\\
        $^{73}$Ge & 7.76$\%$ & 6.52$\%$ (21)\\
        $^{74}$Ge & 36.5$\%$ & 30.75$\%$ (99)\\
        $^{76}$Ge & 7.75$\%$ & 4.97$\%$ (16)\\
        $^{??}$Ge &   -      & 12.73$\%$ (41)\\
          \hline
    \end{tabular}
\end{table}

\begin{table}[h]
\centering
\caption{Summary statistics of the PLE clusters.}

\begin{tabular}{lccc}

\hline
Isotope& Count & Linewidth [MHz] & ZPL Range [GHz] \\
\hline
$^{70}$Ge & 20/58 & 88 $\pm$ 54  & 7.66 \\
$^{72}$Ge & 15/58 & 67 $\pm$ 30  & 4.12 \\
$^{73}$Ge &  5/58 & 181 $\pm$ 36  & 2.52 \\
$^{74}$Ge & 13/58 &  47 $\pm$ 11  & 2.14 \\
$^{??}$Ge & 5/58 & 98 $\pm$ 64&  \\%14.47 \\
\hline
\end{tabular}

\label{tab:iso_summary}
\end{table}

  \putbib[bibs/references]
\end{bibunit}
\clearpage
\section*{Acknowledgments}
This research was primarily supported by  the U.S. Department of Energy, Office of Science; Basic Energy Sciences, Materials Sciences, and Engineering Division (M.W.N.N., B.P., D.D.A., F.J.H., and N.D.). This work was supported by the Laboratory Directed Research and Development (LDRD) Program at Argonne National Laboratory (Project No. 2020-0204), supported by the U.S. Department of Energy, Office of Science, under Contract No. DE-AC02-06CH11357. I.N.H. acknowledges support from the Quantum Leap Challenge Institute for Hybrid Quantum Architectures and Networks (HQAN) (NSF OMA-2016136) and NSF award AM-2240399. A.A.H. acknowledges additional support from  Q-NEXT, a U.S. Department of Energy Office of Science National Quantum Information Science Research Centers under Award Number DE-FOA-0002253. A.W.W. acknowledges support by NSF QuBBE QLCI (NSF OMA-2121044). The spin measurements at JILA was supported by the National Science Foundation (NSF) (2326628 and 2443684). S.S. acknowledges support from the Sloan Research Fellowship.
\clearpage
%\resetlinenumber[1]
\setcounter{figure}{0}
\setcounter{table}{0}
\setcounter{section}{0}
\setcounter{equation}{0}
%\resetlinenumber
\begin{bibunit}[Science]
  
\section{Supplemental Information}
\subsection{SRIM Calculation}

\begin{figure}[H]
    \centering
    \includegraphics[]{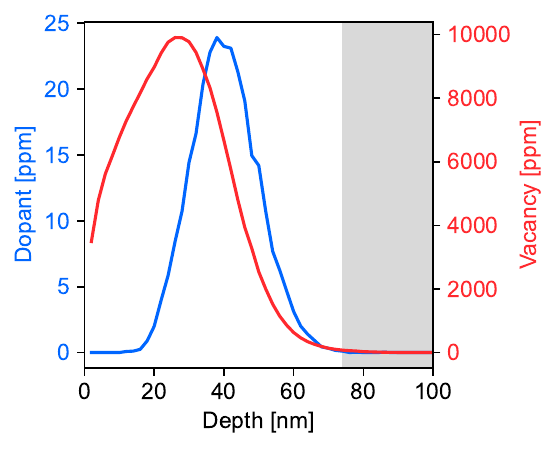}
    %The trim order is left bottom right top.
    \caption{SRIM computation for $^{74}$Ge implanted into diamond at an angle of 7$^\circ$ and 98 keV. The blue line is the dopant profile, and red line is the Vacancy profile. Profiles have been scaled to reflect a $10^{13}$ ions/cm$^2$ dose.}
    \label{sifig:SRIM}
\end{figure}

\subsection{Growth and SIMS}
\begin{figure}[H]
    \centering
    \includegraphics[]{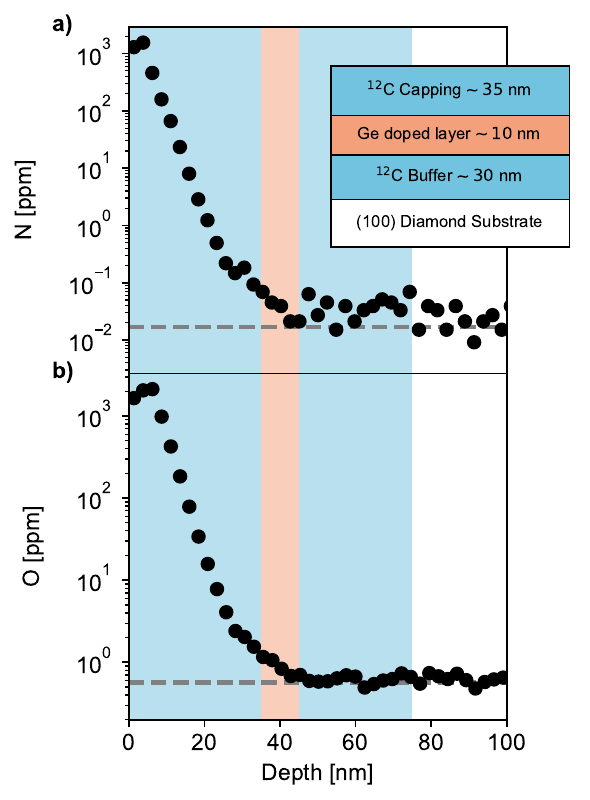}
    %The trim order is left bottom right top.
    \caption{SIMS characterization for a) Nitrogen and b) Oxygen. Detection limits are given with the dashed line at 17 ppb and 570 ppb, respectively. The large surface signal is a SIMS artifact. inset) A schematic illustrating the diamond overgrowth stack with approximate thicknesses.}
    \label{sifig:SI_sims_NO}
\end{figure}

\subsection{Annealing PL Map}
\begin{figure}[H]
    \centering
    \includegraphics[]{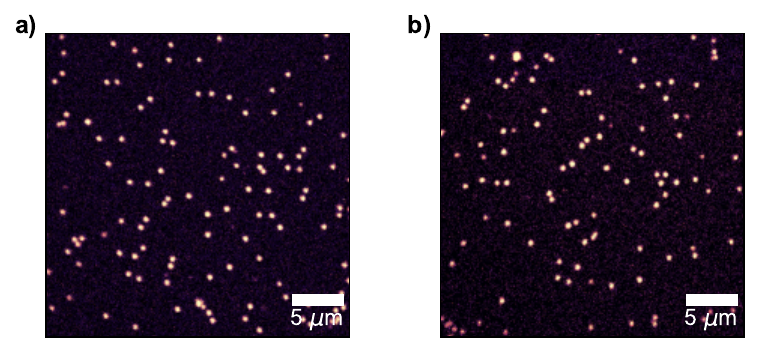}
    %The trim order is left bottom right top.
    \caption{a) PL map after 850 $^\circ$C Anneal. 118 GeVs were observed in an 900 $\mu$m$^2$ area. b) PL map after 1400 $^\circ$C Anneal. 104 GeVs were observed in an 900 $\mu$m$^2$ area. There is no significant change in GeV$^-$ density (Poisson test verification).}
    \label{sifig:SI_Anneal}
\end{figure}

\subsection{PLE}
In this section, examples of the PLE scans are shared. Most of the color centers are stable with some ionization events, but recover immediately: figure \ref{sifig:SI_PLE_96}, \ref{sifig:SI_PLE_104}, \ref{sifig:SI_PLE_163}, \ref{sifig:SI_PLE_194}, \ref{sifig:SI_PLE_222}, \ref{sifig:SI_PLE_133}. The double peaked \ch{GeV-} is given in figure \ref{sifig:SI_PLE_165}. A detailed example of a $^{73}$Ge PLE is given in figure \ref{sifig:SI_PLE_101}. Moreover, other color centers have an instability with bursts of luminescence or irregular lineshapes: figure \ref{sifig:SI_PLE_112}, \ref{sifig:SI_PLE_166}, and \ref{sifig:SI_PLE_297}. 
\subsubsection{Stable PLE}

\begin{figure}[H]
    \centering
    \includegraphics[]{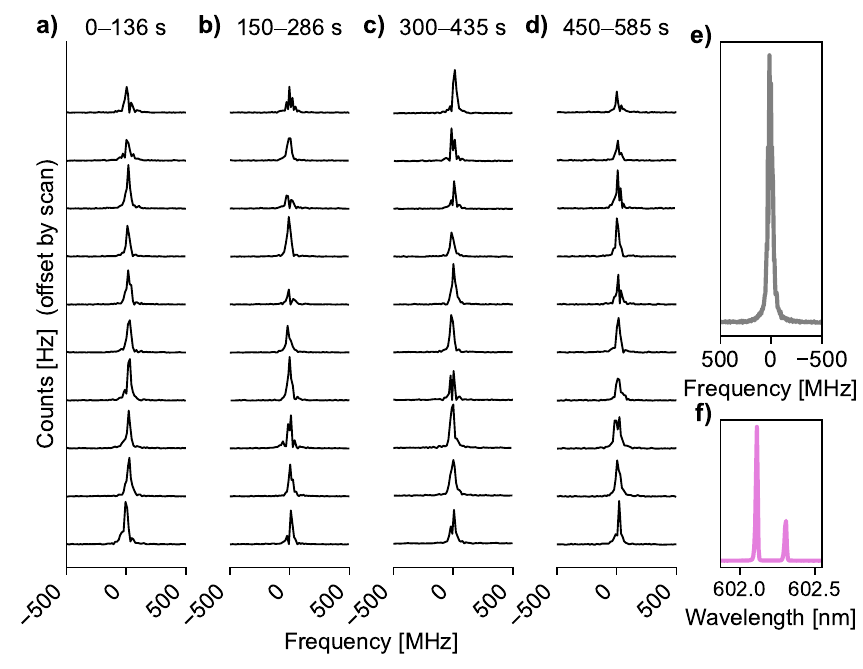}
    %The trim order is left bottom right top.
    \caption{a-e) Time trace of the PLE with a broken axis to highlight the PLE transition. f) Histogram of the PLE. g) PL Spectra. This color center is polarized along the $[1\bar{1}0]$, with a 160 GHz orbital splitting. The average power during PLE is around 4 nW. The fitted linewidth is 47 MHz. This color center is with the $^{70}$Ge isotope.}
    \label{sifig:SI_PLE_96}
\end{figure}

\begin{figure}[H]
    \centering
    \includegraphics[]{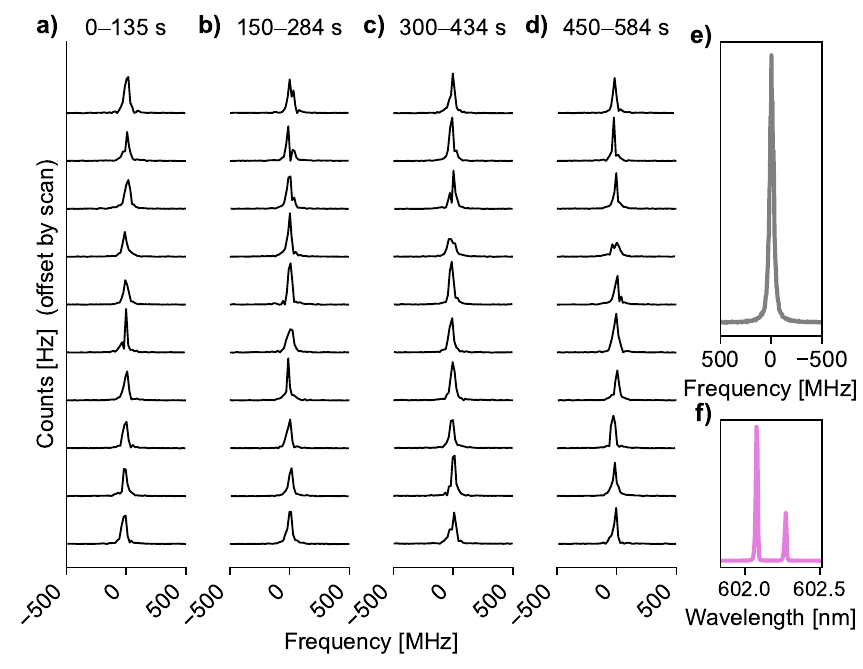}
    %The trim order is left bottom right top.
    \caption{a-e) Time trace of the PLE with a broken axis to highlight the PLE transition. f) Histogram of the PLE. g) PL Spectra. This color center is polarized along the $[1\bar{1}0]$, with a 160 GHz orbital splitting. The average power during PLE is around 4 nW. The fitted linewidth is 45 MHz. This color center is with the $^{70}$Ge isotope.}
    \label{sifig:SI_PLE_104}
\end{figure}

\begin{figure}[H]
    \centering
    \includegraphics[]{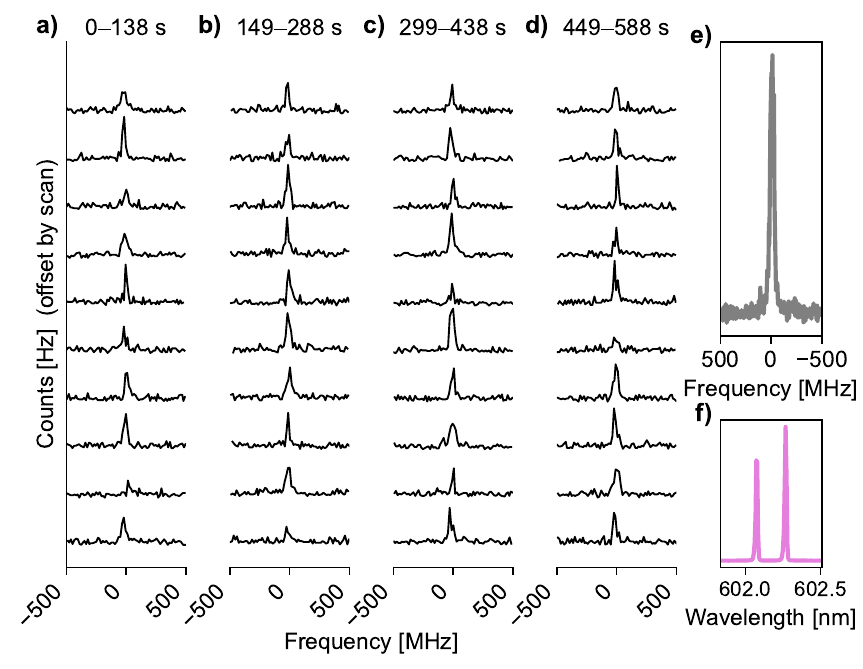}
    %The trim order is left bottom right top.
    \caption{a-e) Time trace of the PLE with a broken axis to highlight the PLE transition. f) Histogram of the PLE. g) PL Spectra. This color center is polarized along the $[110]$, with a 160 GHz orbital splitting. The average power during PLE is around 3.5 nW. The fitted linewidth is 41 MHz. This color center is with the $^{70}$Ge isotope.}
    \label{sifig:SI_PLE_163}
\end{figure}

\begin{figure}[H]
    \centering
    \includegraphics[]{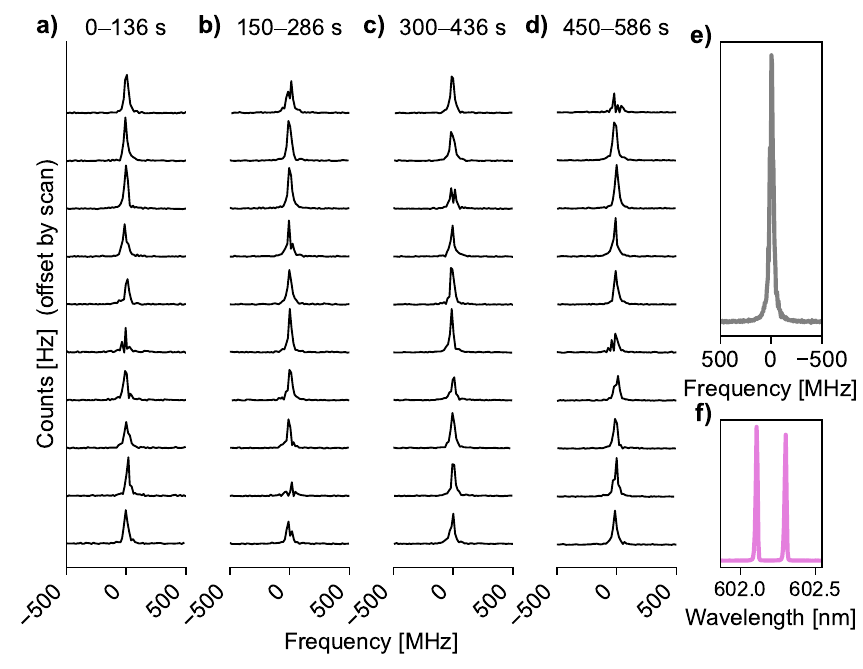}
    %The trim order is left bottom right top.
    \caption{a-e) Time trace of the PLE with a broken axis to highlight the PLE transition. f) Histogram of the PLE. g) PL Spectra. This color center is polarized along the $[110]$, with a 160 GHz orbital splitting. The average power during PLE is around 5 nW. The fitted linewidth is 38 MHz. This color center is with the $^{72}$Ge isotope.}
    \label{sifig:SI_PLE_194}
\end{figure}

\begin{figure}[H]
    \centering
    \includegraphics[]{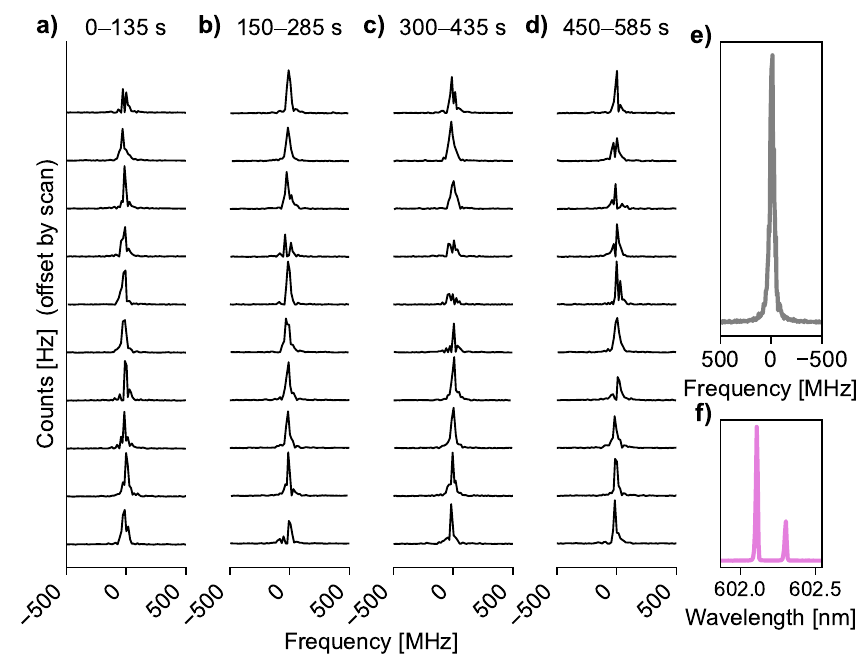}
    %The trim order is left bottom right top.
    \caption{a-e) Time trace of the PLE with a broken axis to highlight the PLE transition. f) Histogram of the PLE. g) PL Spectra. This color center is polarized along the $[1\bar{1}0]$, with a 160 GHz orbital splitting. The average power during PLE is around 4 nW. The fitted linewidth is 42 MHz. This color center is with the $^{72}$Ge isotope.}
    \label{sifig:SI_PLE_222}
\end{figure}

\begin{figure}[H]
    \centering
    \includegraphics[]{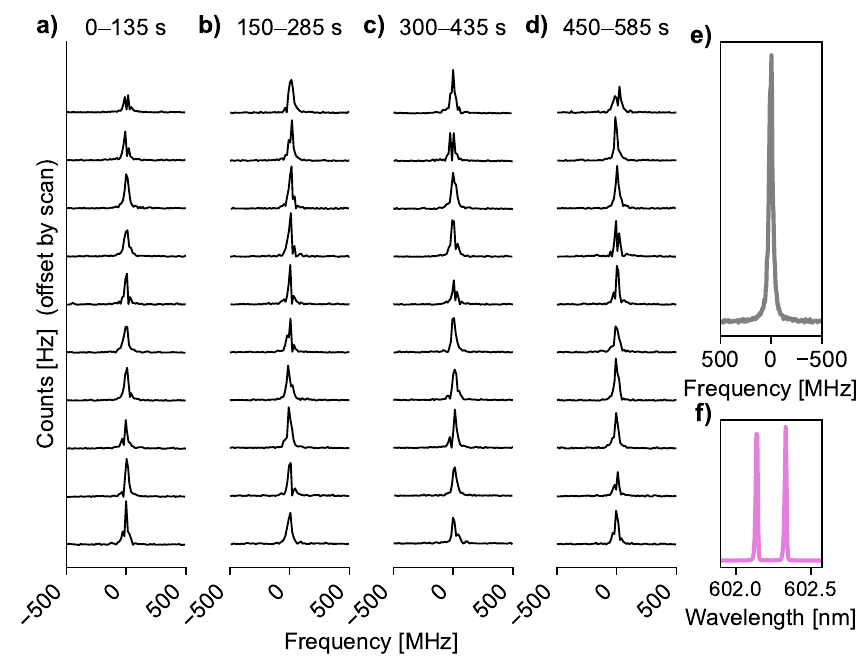}
    %The trim order is left bottom right top.
    \caption{a-e) Time trace of the PLE with a broken axis to highlight the PLE transition. f) Histogram of the PLE. g) PL Spectra. This color center is polarized along the $[110]$, with a 160 GHz orbital splitting. The average power during PLE is around 3 nW. The fitted linewidth is 40 MHz. This color center is with the $^{74}$Ge isotope and is featured in the main text.}
    \label{sifig:SI_PLE_133}
\end{figure}

In the following section PLE time traces are given for selected stable \ch{GeV-}. These color centers recover well against self ionization. Figure \ref{sifig:SI_PLE_96}, \ref{sifig:SI_PLE_104}, \ref{sifig:SI_PLE_163} are from the $^{70}$Ge subset with the first two being polarized along $[1\bar{1}0]$ and the last along $[110]$. Figure \ref{sifig:SI_PLE_194} and \ref{sifig:SI_PLE_222} are from the $^{72}$Ge subset polarized along the  $[110]$ and  $[1\bar{1}0]$. Lastly, Figure \ref{sifig:SI_PLE_133} is the $^{74}$Ge featured in the main text. 

\subsubsection{Double Peak PLE}

\begin{figure}[H]
    \centering
    \includegraphics[]{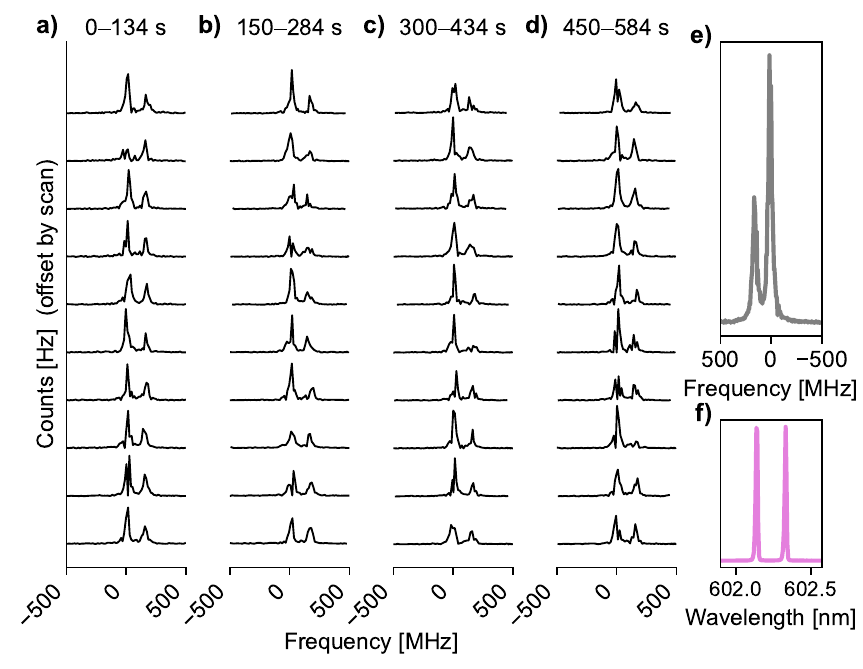}
    %The trim order is left bottom right top.
    \caption{a-e) Time trace of the PLE with a broken axis to highlight the PLE transition. f) Histogram of the PLE. g) PL Spectra. This color center is polarized along the $[110]$, with a 160 GHz orbital splitting. The average power during PLE is around 3 nW. Both peaks are fitted to a linewidth of 42 MHz, their splitting is 152 MHz. This color center is with the $^{74}$Ge isotope.}
    \label{sifig:SI_PLE_165}
\end{figure}

\begin{figure}[H]
    \centering
    \includegraphics[]{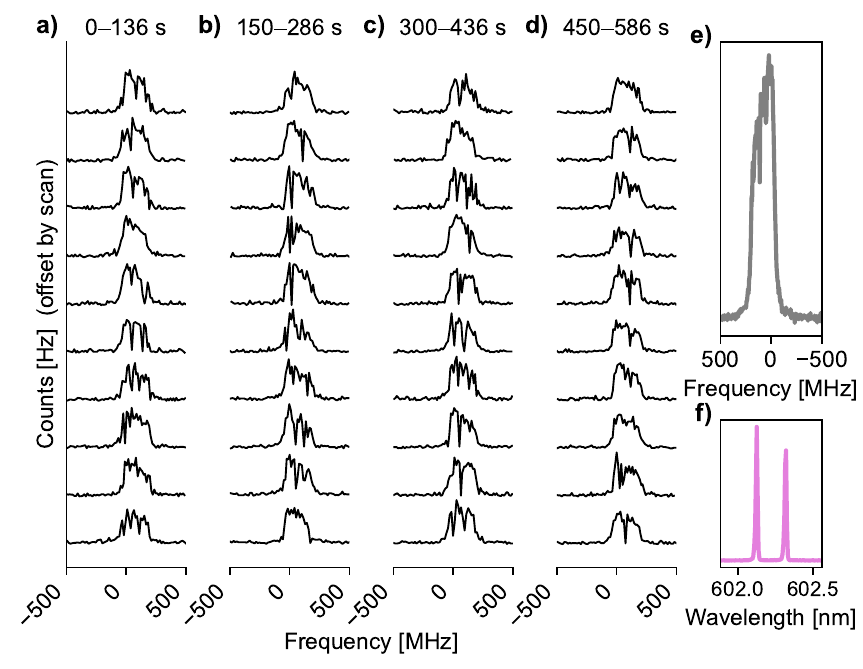}
    %The trim order is left bottom right top.
    \caption{a-e) Time trace of the PLE with a broken axis to highlight the PLE transition. f) Histogram of the PLE. g) PL Spectra. This color center is polarized along the $[110]$, with a 160 GHz orbital splitting. The average power during PLE is around 3 nW. The fitted linewidth is around 208 MHz. This color center is with the $^{73}$Ge isotope.}
    \label{sifig:SI_PLE_101}
\end{figure}

\begin{figure}[H]
    \centering
    \includegraphics[]{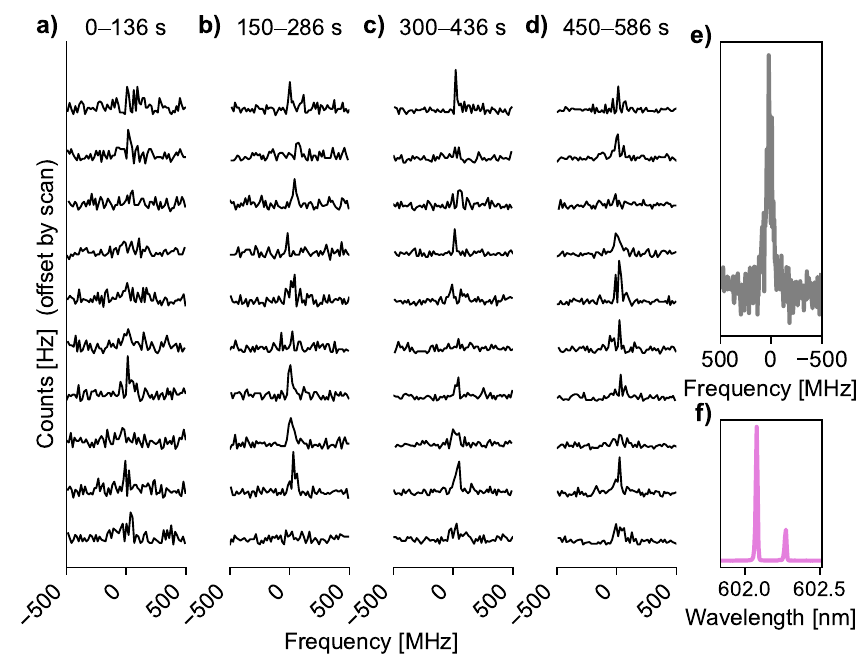}
    %The trim order is left bottom right top.
    \caption{a-e) Time trace of the PLE with a broken axis to highlight the PLE transition. f) Histogram of the PLE. g) PL Spectra. This color center is polarized along the $[1\bar{1}0]$, with a 160 GHz orbital splitting. The average power during PLE is around 3 nW. The fitted linewidth is around 60 MHz. This color is with the $^{70}$Ge isotope.}
    \label{sifig:SI_PLE_112}
\end{figure}

\begin{figure}[H]
    \centering
    \includegraphics[]{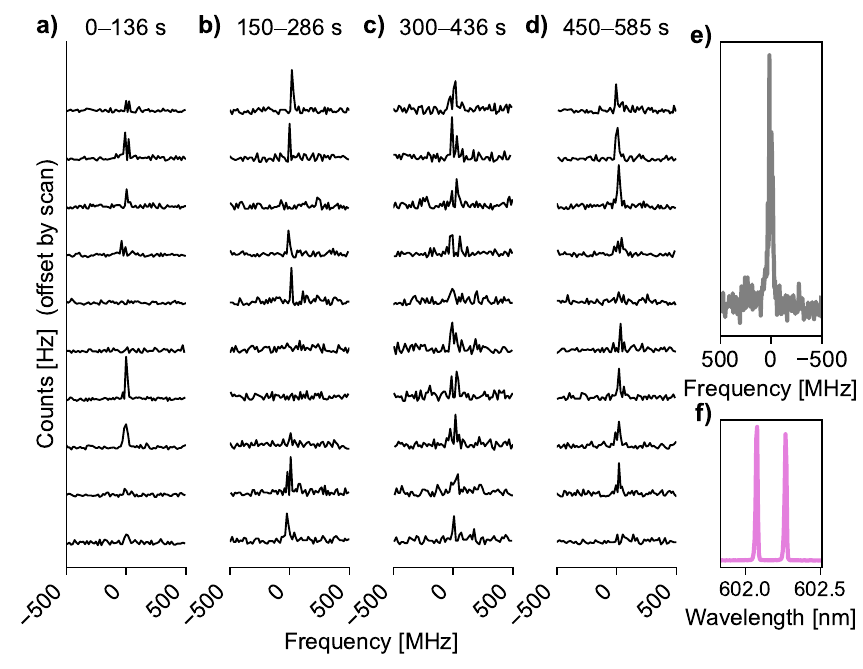}
    %The trim order is left bottom right top.
    \caption{a-e) Time trace of the PLE with a broken axis to highlight the PLE transition. f) Histogram of the PLE. g) PL Spectra. This color center is polarized along the $[110]$, with a 160 GHz orbital splitting. The average power during PLE is around 3.5 nW. The fitted linewidth is around 50 MHz. This color center is with the $^{70}$Ge isotope.}
    \label{sifig:SI_PLE_166}
\end{figure}

\begin{figure}[H]
    \centering
    \includegraphics[]{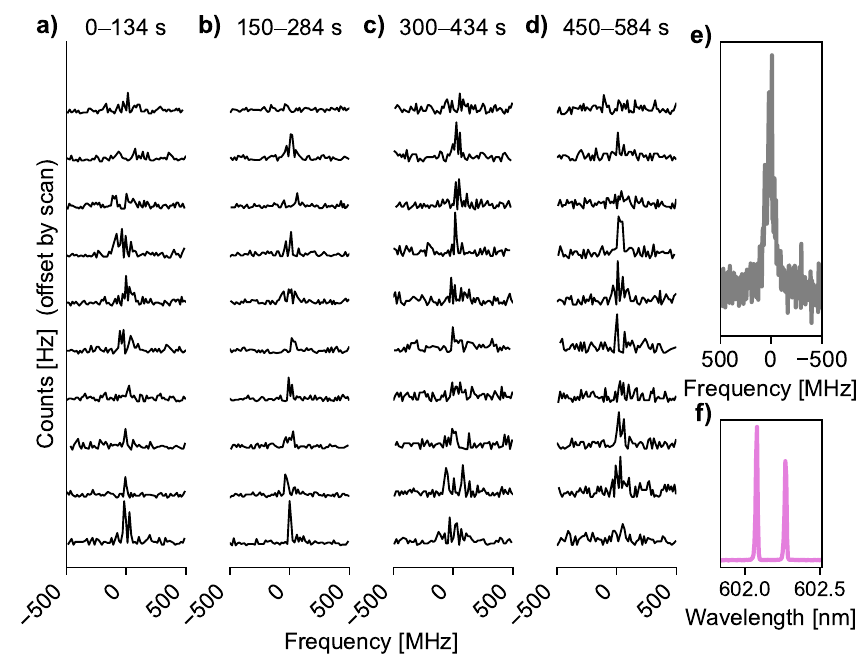}
    %The trim order is left bottom right top.
    \caption{a-e) Time trace of the PLE with a broken axis to highlight the PLE transition. f) Histogram of the PLE. g) PL Spectra. This color center is polarized along the $[110]$, with a 160 GHz orbital splitting. The average power during PLE is around 4 nW. The fitted linewidth is around 50 MHz. This color center is with the $^{70}$Ge isotope.}
    \label{sifig:SI_PLE_297}
\end{figure}

\begin{figure}[H]
    \centering
    \includegraphics[]{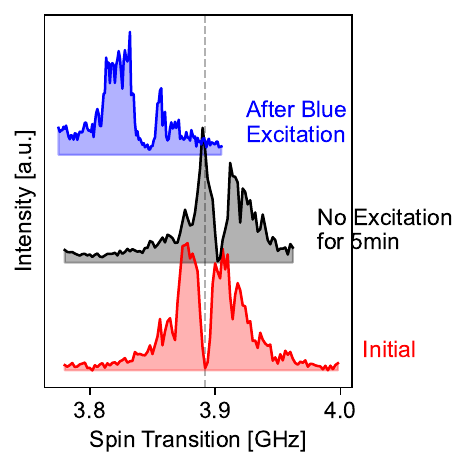}
    %The trim order is left bottom right top.
    \caption{Change of the CPT transition under no excitation, and blue excitation.}
    \label{sifig:SI_CPTChange}
\end{figure}

Figure \ref{sifig:SI_PLE_165} features the double peaked \ch{GeV-}. We estimate the splitting to be 152 MHz. Noticeably from the PLE we donot see a hopping behavior\cite{li_atomic_2024}. Additionally, the self ionization events of the peaks seem slightly correlated. The second order correlation function was not acquired to determine if this is a single emitter or two. 

\subsubsection{$^{73}$Ge PLE}

\subsubsection{Unstable PLE}

\subsection{Unknown Paramagnetic Defect}
\subsubsection{Observation in CPT}

\begin{figure}[H]
    \centering
    \includegraphics[width=\textwidth/2]{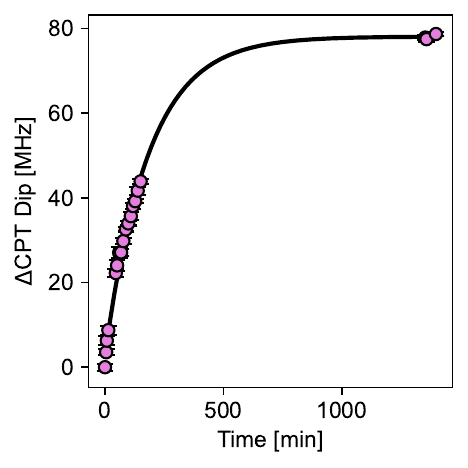}
    %The trim order is left bottom right top.
    \caption{Change of the CPT transition under resonant excitation over the course of 21 hours. We observe a drift around 80 MHz (3 mT). Fitting to an exponential decay, the time constant is around 3 hours.}
    \label{sifig:SI_CPTChangeExpo}
\end{figure}

\subsubsection{Observation in ESR}

\begin{figure}[H]
    \centering
    \includegraphics[]{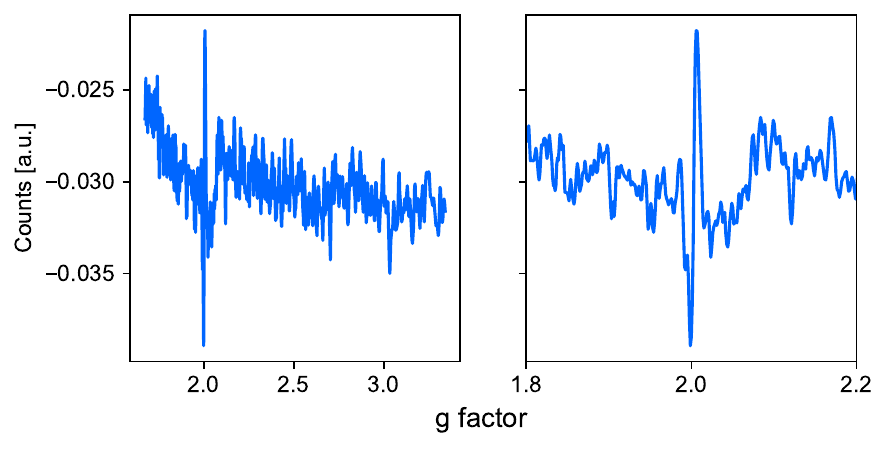}
    %The trim order is left bottom right top.
    \caption{EPR signal at 4.5 K}
    \label{sifig:SI_EPR}
\end{figure}

We performed continuous wave (CW) EPR measurement on the delta doped GeV- sample to gain insight into the GeV-’s spin properties and presence of any other paramagnetic defects. Measurements were conducted using an X-band (9.36 GHz) Bruker ELEXSYS II E500 EPR spectrometer, equipped with a TE102 rectangular EPR resonator. The sample was maintained at cryogenic temperatures around 4.7 K with magnetic field parallel to the diamond sample surface.
Figure 17 shows the measured EPR spectrum over a magnetic field range of 2000-4000 G, using a 1 G B-field modulation. This range corresponds to g-factors from approximately 1.6 to 3.4 and covers the well-known region for defects in diamond\cite{wyk_epr_1994}. As shown in the left panel, only a resonance at g $\approx$ 2 is observed. The resonance has a narrow linewidth of 32 MHz, as shown in the zoomed in plot on the right. This resonance is most likely the associated with GeV- centers containing predominately the even Ge isotope ($^{70}$Ge, $^{72}$Ge, $^{74}$Ge and $^{76}$Ge). No additional resonance were detected. The concentration of any possible co-existing paragenetic defects in the sample is likely below the EPR detection limit.

\subsubsection{Discussion}

We can use a few magneto-static models to estimate the concentration of this defect. To summarize the experimental results: We observe a shift of 80 MHz of the \ch{GeV-} spin transition, see SI figure  \ref{sifig:SI_CPTChangeExpo}. This shift corresponds to a field change of around 3 mT. The applied field in this experiment is 330 mT along the diamond's $\langle 100 \rangle$, we note that the shift is observed for multiple field directions, however due to experimental constraints were unable to determine if the maximum shift depended on applied field direction or intensity. 

We can use Curie's Law in the fully saturated regime to estimate the required density to generate a 3 mT field under an applied 330 mT at 20 mK, see equation \ref{eq:Curie} and \ref{eq:mu}. We take $\mu_B\mu_0 \approx 1.165 \times10^{-20}$ mT/cm$^{3}$. Assuming a magnetic moment of approximately $\mu_B$ for a system of g factor 2 and spin 1/2, we estimate a density of 1500 ppm. This density is very unexpected and does not corroborate with any typical defect that could be created throughout the processing. To recall, we had nominally 17 ppm of Ge at the delta doped layer's peak, a background detection limit of 17 ppb for Nitrogen, and an initial irradiated density of 0.25 ppm of vacancies. 

\begin{figure}[H]
    \centering
    \includegraphics[]{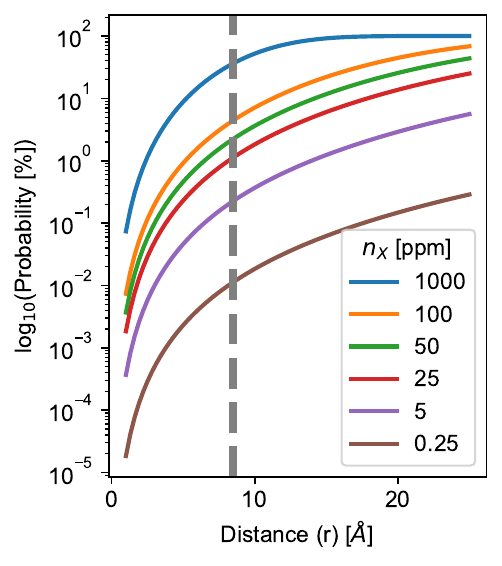}
    %The trim order is left bottom right top.
    \caption{Log plot of probability to find a defect, X, at a distance $\leq$ r for different concentrations of X ($n_X$). Dashed grey line is at 8.5 $\AA$: the distance in which a single defect can produce a 3 mT field.}
    \label{sifig:SI_prob}
\end{figure}

\begin{figure}[H]
    \centering
    \includegraphics[]{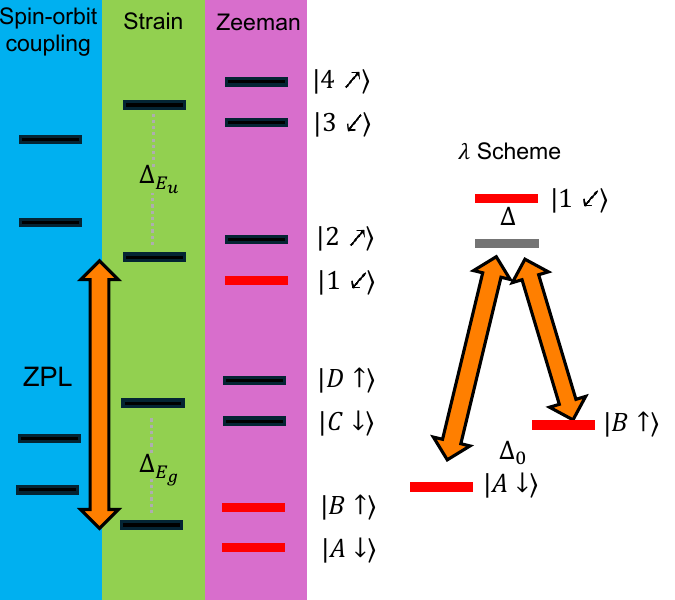}
    %The trim order is left bottom right top.
    \caption{Schematic of the group IV electronic diagram, and the selected levels to form a lambda system for optical control.}
    \label{sifig:SI_lamb}
\end{figure}

Furthermore, we can estimate the probability to find a magnetic defect close enough to our \ch{GeV-} to generate such a field. We use the largest magnetic field from a single electron, see equation \ref{eq:Bdip}, and find that at a distance, $r < 8.5 ~\AA$ the magnetic field, $B ~\geq~ 3$mT. This distance corresponds to about 5 to 6 bond lengths. We then use the probability to find a defect within a radius of a \ch{GeV^-}, see equation \ref{eq:paraProb}. Taking $\rho_D$ to be diamond density, and $n_X$ to be defect concentration we plot probabilities for different concentrations as a function of distance, see figure \ref{sifig:SI_prob}. We find again, that the concentrations required to have this be probable are unreasonable given the material processing and analysis we have done. This suggests the presence of an unknown impurity with a significantly lower density and high magnetic moment or an impurity that preferentially incorporates near the germanium during CVD growth. Per the discussion in the main text, the ionized N$_s$, commonly referred to as the P1 center, is a potential candidate. If it is indeed the P1 center, this poses an interesting caveat with Fermi engineering: the codopants have to be readily ionized and cannot possess paramagnetic properties. If it is the P1 center, a potential solution is to use a higher working temperature, achieved by transverse strain\cite{guo_microwave-based_2023} to leave the P1 in a mixed state. We also suspect it could be a contaminant in the germane gas, or it is from the parent substrate. Further experiments will reuse this gas in the membrane platform to deconvolve these effects.

\begin{equation}
  \delta\textbf{B} \approx \mu_0 \textbf{M} =  \mu_0 \mu n
     \label{eq:Curie}
\end{equation}

\begin{equation}
    \mu = g S \mu_B 
     \label{eq:mu}
\end{equation}

% \begin{equation}
%   \textbf{B}_{dip}(\textbf{r}) = \frac{\mu_0}{4\pi} \left( \frac{3\textbf{r}(\textbf{m}\cdot\textbf{r})}{|r|^5} - \frac{\textbf{m}}{|r|^2} \right)
%      \label{eq:Bdip}
% \end{equation}

\begin{equation}
  \textbf{B}_{dip}(\textbf{r}) = \frac{\mu_0}{4\pi} \frac{2\mu}{|r|^3} 
     \label{eq:Bdip}
\end{equation}

\begin{equation}
    \text{P}( \text{X} < r~| \text{GeV}^-) = 1- \text{exp}(-\frac{4\pi}{3}r^3 \rho_D n_\text{X})
    \label{eq:paraProb}
\end{equation}

\subsection{CPT Details}

\subsubsection{Schematic}

\subsubsection{Fit Details}
To fit the coherent population trapping, we use a Lorentzian dip on a PLE lorentzian. Since we sampled with the zero order of the laser slightly detuned from the spin transition we observe some asymmetry in our CPT scans. To accurately model this, we need to include the contribution from the PLE dome, see equation \ref{eq:CPT01}. However, since we only scanned over a narrowband we can use a Taylor expansion on the PLE dome around the CPT transition: $\Delta = \Delta_0$ and find equation \ref{eq:CPT02}. 

\begin{table}[H]
    \centering
    \begin{tabular}{cl}
        \hline
        \textbf{Variable} & \textbf{Definition} \\
        \hline
        $S(\Delta)$ & Experimental Luminescence\\[6pt]
        $\Delta$ & Two photon detuning of the lasers\\[6pt]
        $\Delta_0$ & Spin Transition\\[6pt]
        $\Delta'$ & Detuning of the PLE dome from the Spin Transition\\[6pt]
        $\gamma_{CPT}$ & Spin Transition Linewidth\\[6pt]
        $\gamma_{PLE}$ & Linewidth of the PLE dome\\[6pt]
        $A$ & PLE Dome amplitude \\[6pt]
        $r$ & ratio of CPT dip to PLE dome\\[6pt]
        $B$ & background counts\\[6pt]
        \hline
    \end{tabular}
    \caption{Definition of variables in equation \ref{eq:CPT01}.}
    \label{tab:variables}
\end{table}

\begin{equation}
    S(\Delta) = \frac{-rA}{1 + 4\left(\frac{\Delta - \Delta_0}{\gamma_{CPT}}\right)^2} + \frac{A}{1 + 4\left(\frac{\Delta - (\Delta_0 - \Delta')}{\gamma_{PLE}}\right)^2} + B
     \label{eq:CPT01}
\end{equation}

\begin{equation}
    \frac{A}{1 + 4\left(\frac{\Delta - (\Delta_0 - \Delta')}{\gamma_{PLE}}\right)^2} \approx C_2(\Delta - \Delta_0)^2 + C_1(\Delta - \Delta_0) + C_0
    \label{eq:CPT02}
\end{equation}

\begin{equation}
    C_2 = -4\gamma_{PLE}^2 A \frac{\gamma_{PLE}^2 - 12\Delta'^2}{\left(4\Delta'^2 + \gamma_{PLE}^2\right)^3}
    \label{eq:C2}
\end{equation}

\begin{equation}
    C_1 = - 8\gamma_{PLE}^2 A \frac{\Delta'}{\left(4\Delta'^2 + \gamma_{PLE}^2\right)^2}
    \label{eq:C1}
\end{equation}

\begin{equation}
    C_0 = A\frac{1}{1 + 4\left(\frac{\Delta'}{\gamma_{PLE}}\right)^2}
    \label{eq:C0}
\end{equation}

  \section*{Supplementary References}
  \putbib[bibs/references]
\end{bibunit}

\end{document}